\documentclass[aps, prl, twocolumn, superscriptaddress, letterpaper, nofootinbib]{revtex4-2}

\usepackage{graphicx}
\usepackage{color}
\usepackage{amssymb}
\usepackage{amsmath}
\usepackage{txfonts}
\usepackage{bm}
\usepackage[colorlinks=true, linkcolor=blue, anchorcolor=black, citecolor=blue, filecolor=black, menucolor=black, runcolor=black, urlcolor=blue]{hyperref}

\newcommand\blfootnote[1]{%
  \begingroup
  \renewcommand\thefootnote{}\footnote{#1}%
  \addtocounter{footnote}{-1}%
  \endgroup
}

\begin{document}

\title{Three-dimensional valley-contrasting sound}

\author{Haoran Xue*$\dagger$}
\affiliation{Department of Physics, The Chinese University of Hong Kong, Shatin, Hong Kong SAR, China}

\author{Yong Ge$\dagger$}
\affiliation{Research Center of Fluid Machinery Engineering and Technology, School of Physics and Electronic Engineering, Jiangsu University, Zhenjiang 212013, China}

\author{Zheyu Cheng}
\affiliation{Division of Physics and Applied Physics, School of Physical and Mathematical Sciences, Nanyang Technological University,
Singapore 637371, Singapore}

\author{Yi-jun Guan}
\affiliation{Research Center of Fluid Machinery Engineering and Technology, School of Physics and Electronic Engineering, Jiangsu University, Zhenjiang 212013, China}

\author{Jiaojiao Zhu}
\address{Research Laboratory for Quantum Materials, Singapore University of Technology and Design, Singapore 487372, Singapore}

\author{Hong-yu Zou}
\affiliation{Research Center of Fluid Machinery Engineering and Technology, School of Physics and Electronic Engineering, Jiangsu University, Zhenjiang 212013, China}

\author{Shou-qi Yuan}
\affiliation{Research Center of Fluid Machinery Engineering and Technology, School of Physics and Electronic Engineering, Jiangsu University, Zhenjiang 212013, China}

\author{Shengyuan A. Yang}
\address{Research Laboratory for Quantum Materials, IAPME, University of Macau, Macau SAR, China}

\author{Hong-xiang Sun*}
\affiliation{Research Center of Fluid Machinery Engineering and Technology, School of Physics and Electronic Engineering, Jiangsu University, Zhenjiang 212013, China}
\affiliation{State Key Laboratory of Acoustics, Institute of Acoustics, Chinese Academy of Sciences, Beijing 100190, China.}

\author{Yidong Chong*}
\affiliation{Division of Physics and Applied Physics, School of Physical and Mathematical Sciences, Nanyang Technological University,
Singapore 637371, Singapore}
\affiliation{Centre for Disruptive Photonic Technologies, Nanyang Technological University, Singapore 637371, Singapore}

\author{Baile Zhang*}
\affiliation{Division of Physics and Applied Physics, School of Physical and Mathematical Sciences, Nanyang Technological University,
Singapore 637371, Singapore}
\affiliation{Centre for Disruptive Photonic Technologies, Nanyang Technological University, Singapore 637371, Singapore}

\maketitle
\textbf{Spin and valley are two fundamental properties of electrons in crystals. The similarity between them is well understood in valley-contrasting physics established decades ago in two-dimensional (2D) materials like graphene -- with broken inversion symmetry, the two valleys in graphene exhibit opposite orbital magnetic moments, similar to the spin-1/2 behaviors of electrons, and opposite Berry curvature that leads to a half topological charge. However, valley-contrasting physics has never been explored in three-dimensional (3D) crystals. Here, we develop a 3D acoustic crystal exhibiting 3D valley-contrasting physics. Unlike spin that is fundamentally binary, valley in 3D can take six different values, each carrying a vortex in a distinct direction. The topological valley transport is generalized from the edge states of 2D materials to the surface states of 3D materials, with interesting features including robust propagation, topological refraction, and valley-cavity localization. Our results open a new route for wave manipulation in 3D space.}

\noindent\textbf{Introduction}\blfootnote{*Corresponding author. Email: haoranxue@cuhk.edu.hk (H.X.); jsdxshx@ujs.edu.cn (H.S.); yidong@ntu.edu.sg (Y.C.); blzhang@ntu.edu.sg (B.Z.). $\dagger$These authors contributed equally to this work.}

In a crystalline material, a ``valley'' refers to a set of local energy extrema in the band energies~\cite{schaibley2016}, notably observed in 2D graphene-like materials with broken inversion symmetry, showcasing two distinct valleys. These valleys possess contrasting band geometric properties~\cite{xiao2007, yao2008}: firstly, exhibiting an opposite orbital magnetic moment that circulates either clockwise or counterclockwise within each valley, akin to the spin-1/2 behavior of electrons. Secondly, they feature oppositely distributed Berry curvature, whereby the integration within each valley is quantized to a half topological charge characterized by a half Chern number. These contrasting band geometric properties give rise to striking phenomena such as the valley Hall effect~\cite{mak2014, gorbachev2014} and topological valley transport \cite{ju2015, lu2017, gao2018}. The emerging field of valleytronics is based on using the valley degree of freedom for information storage and manipulation~\cite{schaibley2016}, akin to the role of spin in spintronics. Notably, the valley degree of freedom is not limited to electronic materials, but can also occur in classical waves within photonic or acoustic crystals~\cite{lu2016, ma2016, lu2017, dong2017, gao2018, yan2018, xue2021, xue2022}. Owing to its simple design and high performance, valley-based waveguides have emerged as a leading candidate for topologically protected waveguiding of classical waves, with promising applications in on-chip signal processing~\cite{yan2018}, topological lasing~\cite{zeng2020}, and quantum optics~\cite{mehrabad2020}.

The study of valley-contrasting physics originated in and has thus far been restricted to 2D crystals.  However, the basic notion---band geometric properties taking on sharply-contrasting values around different points in momentum space~\cite{xiao2007}---has no fundamental restriction to 2D, and should therefore be achievable in 3D. In this work, we design and implement a 3D acoustic crystal with valley-contrasting physics. A key feature of this crystal is that the band geometric properties have 3D vectorial characteristics. Instead of the two valleys of 2D graphene-like lattices, the 3D lattice hosts six valleys, divided into three pairs; for each pair, the Berry curvature and orbital magnetic moment point along a unique axis, and take on opposite values within the pair. The integration of Berry curvature in each valley, while still quantized to a half Chern number, takes a 3D direction accordingly. The 3D valley states also exhibit properties qualitatively distinct from the 2D case. In the bulk, they form vortices oriented along different 3D directions (unlike valley states in 2D, which can only circulate clockwise or counterclockwise). At the 2D interface between two 3D valley acoustic crystals with opposite band masses, there arise robust directional valley kink states, which generalize the kink states that occur along one-dimensional (1D) boundaries of 2D valley materials.  Using a range of acoustic measurements, we experimentally verify the properties of the acoustic crystal's 3D valley states.

\begin{figure*}
  \centering
  \includegraphics[width=\textwidth]{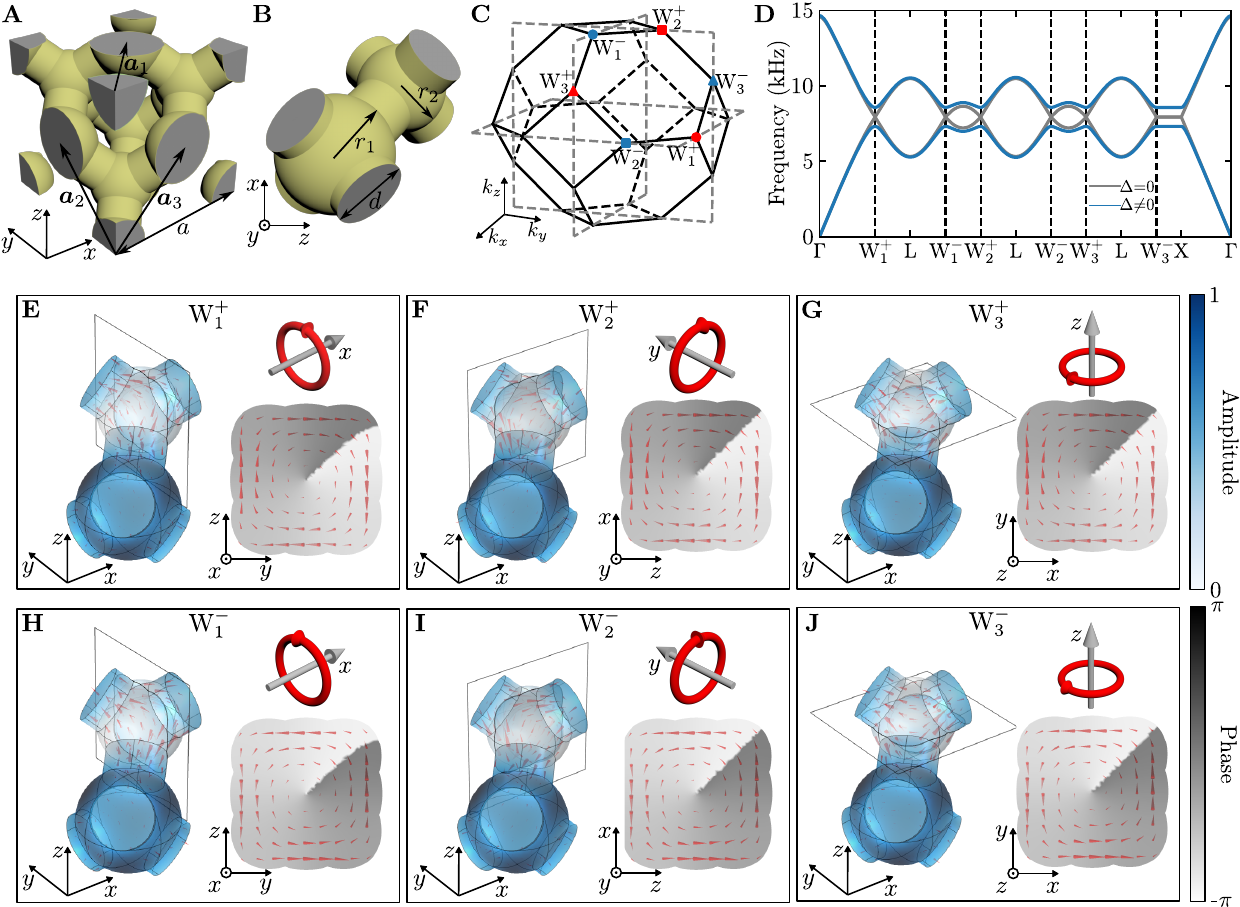}
  \caption{\textbf{Vectorial valley-contrasting physics in a 3D acoustic crystal.} (\textbf{A} and \textbf{B}) Cubic cell (A) and unit cell (B) of the valley acoustic crystal, with the lattice vectors $\bm{a_1}=(a/2,a/2,0)$, $\bm{a_2}=(0,a/2,a/2)$ and $\bm{a_3}=(a/2,0,a/2)$. The grey region is filled with air and the yellow shells represent rigid walls. (\textbf{C}) First Brillouin zone, with the six valleys denoted by colored markers. Grey dashes indicate the nodal lines for $\Delta=0$. (\textbf{D}) Bulk dispersion of the acoustic crystal. Grey and blue curves correspond to the cases where $\Delta=0$ and $\Delta=1$ mm, respectively. (\textbf{E} to \textbf{J}) Bloch modes for the lowest band at the six valleys. The blue and grey colors denote the amplitude and phase of the pressure field, and the red arrows indicate the power flow. The subplots at the upper-right corner indicate the phase winding patterns of the modes.}
  \label{fig1}
\end{figure*}

\noindent\textbf{Results}

\noindent\textbf{Acoustic crystal design}

In 2D, valleys with contrasting band geometric properties are commonly achieved by gapping out Dirac cones. Dirac cones at different points in the Brillouin zone (BZ) map onto each other under time reversal, and hence carry opposite Berry curvature and orbital magnetic moment. To generalize valley-contrasting physics to 3D, we seek a 3D bandstructure with two-fold band degeneracies that can be gapped out. Weyl points, commonly regarded as the 3D generalization of 2D Dirac points, are unsuitable since they cannot be gapped out by perturbations~\cite{wan2011}. We instead seek out a lattice whose bandstructure contains nodal lines (i.e., degeneracies occurring along lines rather than points in momentum space~\cite{fang2016}). Specifically, we identify nodal chains as being promising for producing inhomogeneous band geometric features (Fig.~S1).  A nodal chain can occur in a diamond lattice with nearest-neighbor coupled $s$ orbitals; this lattice also serves as a 3D analog of graphene, hosting many 3D generalizations of 2D phenomena including topological insulators~\cite{kane2005, fu2007}, higher-order topological insulators~\cite{ezawa2018, xue2019}, and pseudo-Landau levels~\cite{guinea2010, rachel2016, cheng2024}. 

Based on these considerations, we design a 3D acoustic crystal based on a 3D tight-binding diamond lattice with on-site detuning (see Supplementary text S1). The cubic cell, shown in Fig.~\ref{fig1}A, consists of hollow spherical cavities connected by cylindrical tubes. In each unit cell, there are two cavities with radii $r_1$ and $r_2$, respectively (Fig.~\ref{fig1}B). When $\Delta\equiv r_1-r_2=0$, the crystal has inversion symmetry and hosts nodal chains, as plotted with grey dashes in Fig.~\ref{fig1}C. For $\Delta \ne 0$, inversion symmetry is broken, which opens a full bandgap (Fig.~\ref{fig1}D); as shown below, this leads to 3D valley-contrasting physics.

\begin{figure*}
  \centering
  \includegraphics[width=\textwidth]{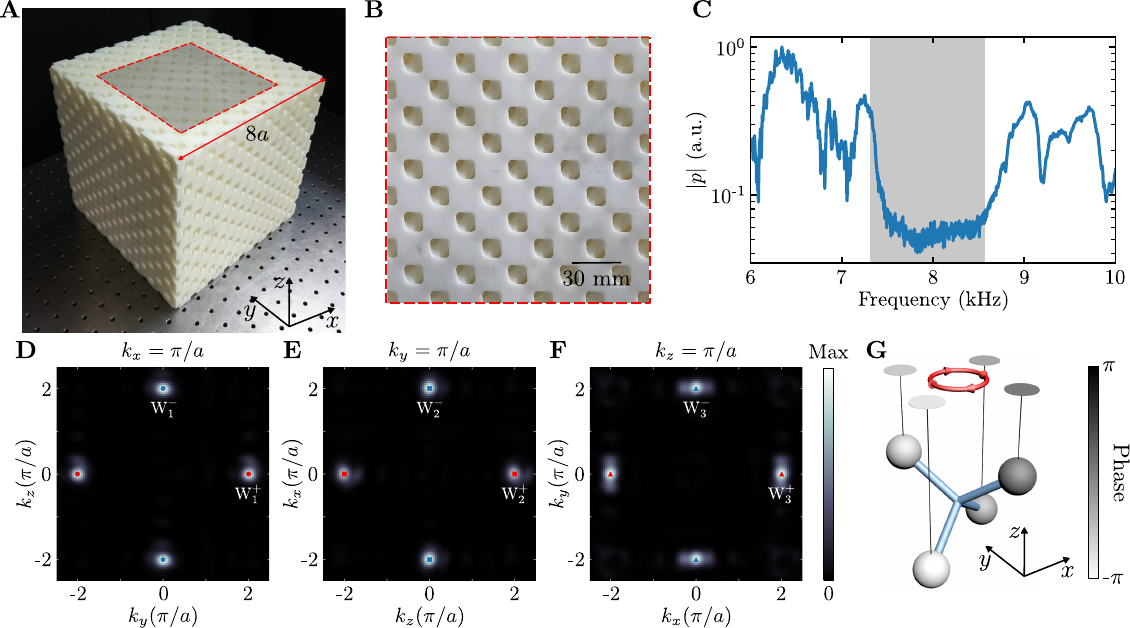}
  \caption{\textbf{Characterization of the bulk valley vortex states.} (\textbf{A}) Photograph of the sample, which consists of eight cubic cells in each direction. (\textbf{B}) A magnified picture displaying the details of the upper surface.  (\textbf{C}) Measured spectrum at a bulk site. The grey region indicates the simulated bandgap. (\textbf{D} to \textbf{F}), Fourier spectra of the acoustic fields at 7300 Hz excited by a point-like source placed at the center. We adopt three 2D cuts in the 3D momentum space that contain the six valleys.  (\textbf{G}) A simplified illustration of the Bloch modes at $\text{W}_3^+$ valley. The four balls denote four neighboring cavities, with their colors representing the phases.}
  \label{fig2}
\end{figure*}

The 3D valleys are centered at the following six momenta (see Fig.~\ref{fig1}C):
\begin{equation}
\begin{aligned}
\text{W}_1^{\pm}&=(\pm\pi/a,2\pi/a,0),\\
\text{W}_2^{\pm}&=(0,\pm\pi/a,2\pi/a),\\
\text{W}_3^{\pm}&=(2\pi/a,0,\pm\pi/a),
\end{aligned}
\end{equation}
where $a=31.5$ mm is the side length of the cubic cell (Fig.~\ref{fig1}A). The subscripts $1,2,3$ denote the three valley pairs and the superscripts $\pm$ denote the two valleys within each pair, which are related by time-reversal symmetry. Within these valleys, the effective Hamiltonians are (Supplementary text S2):
\begin{align}
\mathcal{H}_{\text{W}_1^{\pm}}(\bf{q})&=\sqrt{2}v(\mp\sigma_xq_y+\sigma_yq_z)+\sigma_zm,\label{Heff-W1}\\
\mathcal{H}_{\text{W}_2^{\pm}}(\bf{q})&=\sqrt{2}v(\mp\sigma_xq_z+\sigma_yq_x)+\sigma_zm,\label{Heff-W2}\\
\mathcal{H}_{\text{W}_3^{\pm}}(\bf{q})&=\sqrt{2}v(\mp\sigma_xq_x+\sigma_yq_y)+\sigma_zm.\label{Heff-W3}
\end{align}
Here, ${\bf{q}}=(q_x,q_y,q_z)$ is the momentum relative to each valley center, $v$ is the group velocity, $\omega_0^2=(\omega_1^2+\omega_2^2)/2$ ($\omega_{1,2}$ are the angular frequencies of the lowest two bands at the six valleys), $\sigma_{x,y,z}$ are Pauli matrices, $\sigma_0$ is an identity matrix, and $m$ is the mass term induced by a nonzero $\Delta$.  Each pair of valleys is governed by massive Dirac Hamiltonians of opposite chirality, and hence opposite (and nontrivial) Berry curvature. Different pairs of valley Hamiltonians are distinguished by different combinations of two out of three momentum components.

The valley Bloch modes also carry nonzero orbital magnetic moment. This can be seen by mapping out the phase winding and power flow of the acoustic wave, as shown in Fig.~\ref{fig1}, E--J.  As we might expect from the arrangement of 3D valleys, the circulations for different valley pairs lie in different 2D planes ($yz$, $zx$ or $xy$), with opposite circulations for the two valleys in each pair.

\begin{figure*}
  \centering
  \includegraphics[width=\textwidth]{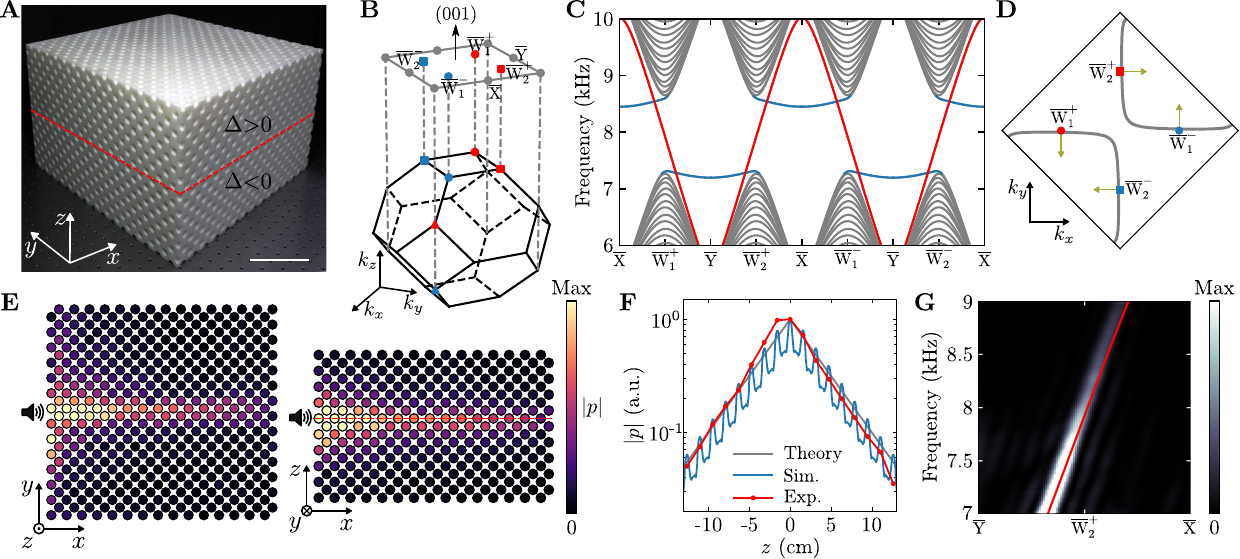}
  \caption{\textbf{Observation of valley kink states.} (\textbf{A}) Photograph of the sample, which consists of two domains with opposite structural parameter $\Delta$. The red dashes denote the 2D interface. (\textbf{B}) (001) surface Brillouin zone. The projections of $\text{W}_1^{\pm}$ and $\text{W}_2^{\pm}$ valleys are labelled by $\overline{\text{W}}_1^{\pm}$ and $\overline{\text{W}}_2^{\pm}$, respectively. (\textbf{C}) Simulated dispersion for the structure in (A), with periodic boundary condition imposed along $x$ and $y$ directions and $z$ direction terminated by hard walls. The red and blue bands correspond to states localized at the interface and the outer boundaries, respectively, while the grey bands are bulk bands. (\textbf{D}) Simulated eigenfrequency contour at 8000 Hz (grey curves) in the surface Brillouin zone. The arrows indicate the propagation directions of the kink states at the valleys. (\textbf{E}) Measured acoustic field distribution (absolute value of sound pressure) at 8000 Hz on the interface (left) and on the middle $xz$ plane (left). The source is denoted by the speaker icon and the interface is highlighted by a red dashed line. (\textbf{F}) The decay profile of the valley kink states. The grey, blue and red curves correspond to theoretical, simulated and measured results, respectively. (\textbf{G}) Comparison between the measured (colormap) and simulated (red line) dispersion of the valley kink state.}
  \label{fig3}
\end{figure*}

\noindent\textbf{Bulk property characterization}

o experimentally probe the bulk properties of this 3D acoustic crystal, we fabricate a sample of size $8a\times 8a\times 8a$, as shown in Fig.~\ref{fig2}, A and B. A point-like sound source is placed at the center of the sample to excite the acoustic fields~\cite{SM}. The spectrum at a bulk site clearly reveals the existence of a bandgap, whose frequency range matches well with the simulation (Fig.~\ref{fig2}C). To further characterize the valley states, we measure the acoustic field distribution inside the whole sample and Fourier transform it to obtain the momentum space distribution. As plotted in Fig.~\ref{fig2}, D-F, the Fourier intensity distributions at 7300 Hz (near the lower band edge) in certain 2D momentum planes that contain the six valleys exhibit high weights around the valleys, showing that there are indeed Bloch modes at the valleys. Moreover, we can recover more information on the Bloch modes by inspecting the Fourier intensities at different sublattices~\cite{peri2020}. At the lower band edge (i.e., 7300 Hz), we find that, around the six valleys, the Fourier intensity in the cavity with radius $r_1$ is much higher than in the one with radius $r_2$, which is consistent with the numerical modes shown in Fig.~\ref{fig1}, E-J. The vortex nature can also be revealed by looking at a simplified Bloch mode pattern of a certain valley. For example, Fig.~\ref{fig2}G depicts the experimental Bloch mode at $\text{W}_3^+$ by plotting the phases of four neighboring cavities with radius $r_1$ (note the Fourier intensity in the cavity of radius $r_2$ is negligible), which shows a clockwise phase winding in the $xy$ plane.

\begin{figure*}
  \centering
  \includegraphics[width=0.85\textwidth]{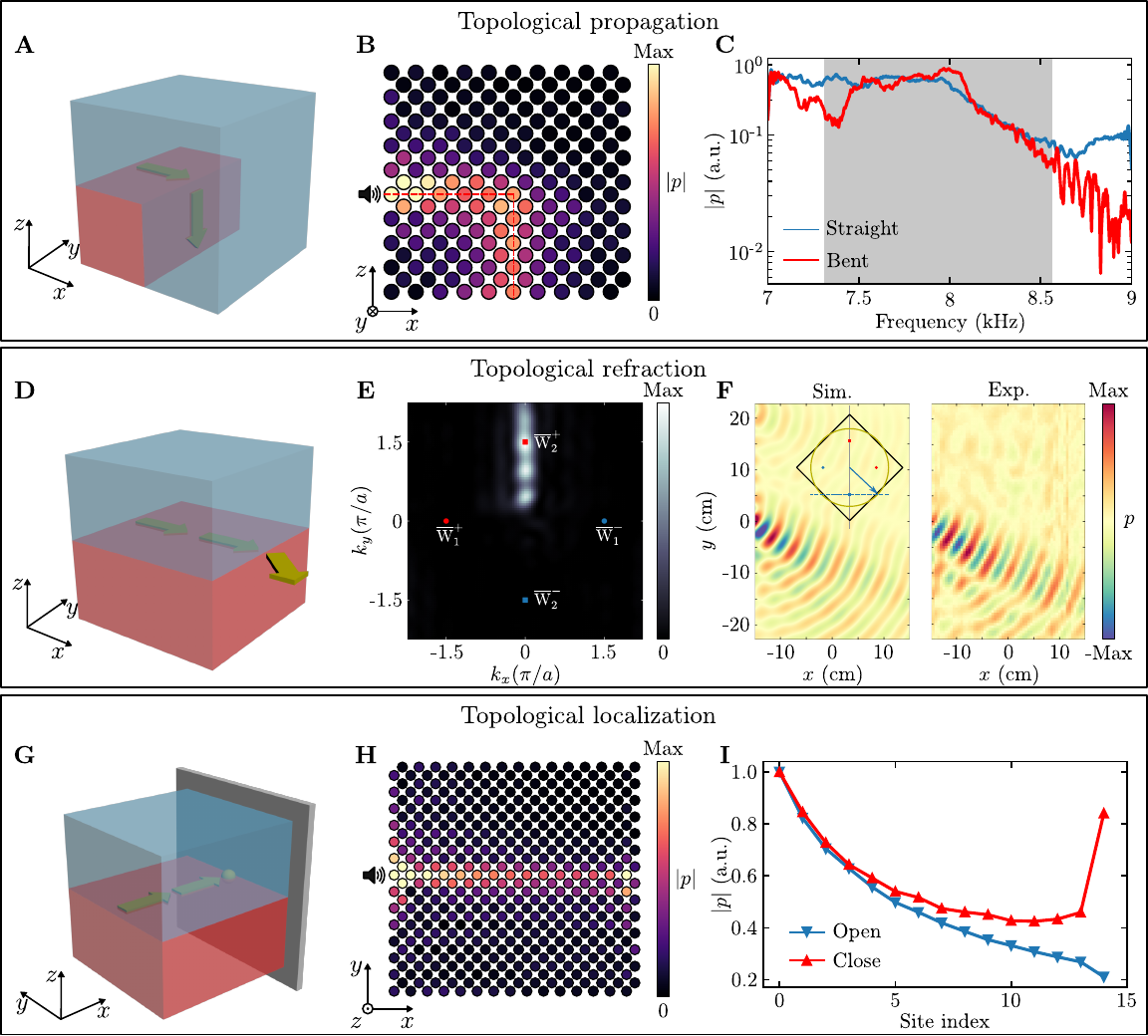}
  \caption{\textbf{Topological properties of valley kink states.} (\textbf{A}) Schematic of the topological propagation of valley kink states through a $90 ^{\circ}$ bend. (\textbf{B}) Measured acoustic field distribution at 8000 Hz on the middle $xz$ plane. The source is denoted by the speaker icon and the interface is highlighted by a red dashed line. (\textbf{C}) Measured transmission spectra of valley kink states for straight (blue curve) and bent (red curve) interfaces. The grey region indicates the bandgap. (\textbf{D}) Schematic of the topological refraction of valley kink states into free space. (\textbf{E}) Fourier spectrum of the acoustic field (at 8000 Hz) at the interface. (\textbf{F}) Simulated (left) and measured (right) acoustic field distribution at 8000 Hz in free space in the $xy$ plane (at the same height as the interface). The inset shows the phase matching diagram for the outcoupling process. (\textbf{G}) Schematic of the topological localization of valley kink states at the interface between the sample and a hard wall. (\textbf{H}) Measured acoustic field distribution at 8000 Hz on the interface. (\textbf{I}) Measured acoustic field distributions at 8000 Hz along the $x$-directional middle line on the interface under open (blue curve) and closed (red curve) boundaries at the outcoupling surface.}
  \label{fig4}
\end{figure*}

\noindent\textbf{Valley kink states}

The 3D valley acoustic crystal hosts 2D valley kink states at interfaces between distinct domains, induced by the valley-contrasting Berry curvature in the bulk (Supplementary text S3). Consider a heterostructure formed by two crystals with opposite $\Delta$, as shown in Fig.~\ref{fig3}A. Along the (001) interface, the $\text{W}_1^{\pm}$ and $\text{W}_2^{\pm}$ valleys are projected to different points in the surface BZ, whereas $\text{W}_3^{\pm}$ project to the same point (see Fig.~\ref{fig3}B). For the $\text{W}_1^+$ valley, we can define a valley Chern number at each $q_{x}$ as
\begin{equation}
C_{\text{W}_1^+}(q_x)=\frac{1}{2\pi}\int_{S_{\text{W}_1^+}}\bm{\Omega}\cdot \text{d}\bm{S},
\end{equation}
where $S_{\text{W}_1^+}$ is a planar region around $\text{W}_1^+$ normal to $q_x$. For the lower band, we find $C_{\text{W}_1^+}(q_x)=\frac{1}{2}\text{sgn}(\Delta)$.  Hence, the valley Chern numbers differ by $\Delta C_{\text{W}_1^+}=-1$ between the lower and upper domains, implying that there is one gapless chiral mode (for each $q_x$) propagating in the $-y$ direction. The kink states for the other valleys can be analyzed similarly, yielding $\Delta C_{\text{W}_1^{\pm}}(q_x)=\mp1$ and $\Delta C_{\text{W}_2^{\pm}}(q_y)=\mp1$. Therefore, the kink states of each valley form a chiral-sheet dispersion, with valley-locked directional propagation. These predictions are confirmed by numerical calculations of the interfacial dispersion relations (Fig.~\ref{fig3}, C and D). We note that the above analysis is based on the effective Hamiltonians that assume small perturbations. Under a large inversion-symmetry breaking, the kink states may only occupy part of the bandgap and exhibit less robustness.

To probe these kink states experimentally, we place a speaker on the left side of the acoustic crystal and measure the sound pressure at different points within the sample. The results show that the sound is localized around the interface and propagates directionally along $x$ (Fig.~\ref{fig3}E). The sound pressure decays exponentially along the interface, consistent with the numerical and theoretical predictions (Fig.~\ref{fig3}F). Along $y$ direction, propagation is also suppressed due to the flat dispersion and the spreading of the sound wave is determined by the width of the excitation. Furthermore, the excited states, which propagate rightwards, should come from the $\text{W}_2^+$ valley and have a linear dispersion (see Fig.~\ref{fig3}, C and D). These properties are verified by the measured dispersion shown in Fig.~\ref{fig3}G, obtained from Fourier transform of the acoustic field at the interface.

In 2D valley materials, kink states are known to exhibit many intriguing behaviors tied to valley conservation~\cite{xue2021}.  These properties generalize to the kink states of the present 3D acoustic crystal (Fig.~\ref{fig4}). For instance, 2D kink states can pass through sharp bends with neglectable reflection~\cite{lu2017}. To study this in 3D, we construct a sample with an interface featuring a $90 ^{\circ}$ bend (Fig.~\ref{fig4}A). Through field imaging and transmission measurement, we find that the kink states indeed go around this path bend without noticeable scattering (Fig.~\ref{fig4}, B and C). It is worth noting that the kink states are not robust against random disorders that cause intervalley scattering, as demonstrated in previous 2D valley photonic crystals~\cite{gao2018, rosiek2023}.

It is also possible for the valley degree of freedom to be conserved at an interface between the sample and free space, leading to the phenomenon of topological refraction~\cite{ma2016, gao2018}. In our 3D system, this occurs when the outcoupling plane is normal to the propagation direction of the kink states (Fig.~\ref{fig4}D). In the depicted configuration, the kink states of $\text{W}_2^+$ refract perfectly into free space, with reflection forbidden due to the suppression of intervalley scattering. This is evident from the Fourier spectrum of the acoustic field inside the sample, which shows negligible wave amplitudes around the other valleys (Fig.~\ref{fig4}E). We can also identify the directionality of the wave refracted into free space (Fig.~\ref{fig4}F), which agrees with the prediction based on phase-matching (Fig.~\ref{fig4}F, inset). 

\begin{figure*}
  \centering
  \includegraphics[width=0.85\textwidth]{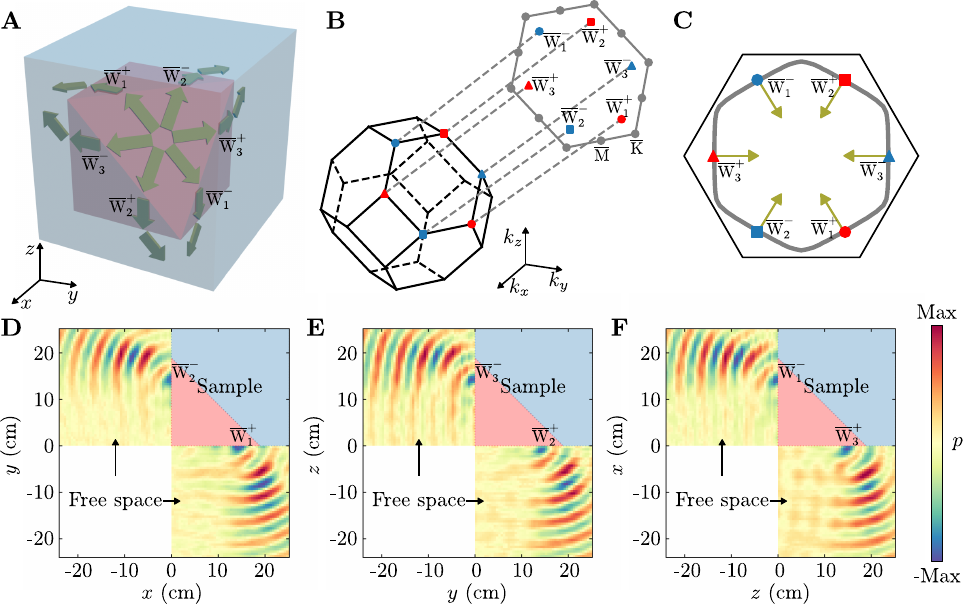}
  \caption{\textbf{3D valley-dependent beam splitting utilizing all six valleys.} (\textbf{A}) Schematic of the 3D valley-dependent beam splitting. (\textbf{B}) (111) surface Brillouin zone. The projections of $\text{W}_1^\pm$, $\text{W}_2^\pm$ and $\text{W}_3^\pm$ valleys are labelled by $\overline{\text{W}}_1^\pm$, $\overline{\text{W}}_2^\pm$ and $\overline{\text{W}}_3^\pm$, respectively. (\textbf{C}) Simulated eigenfrequency contour for the (111) interface at 8000 Hz (grey curve) in the surface Brillouin zone. The arrows indicate the propagation directions of the kink states at the valleys. (\textbf{D} to \textbf{F}) Measured acoustic field distributions at 8000 Hz in the corresponding planes for the outcoupling waves of the six valleys.}
  \label{fig5}
\end{figure*}

The topological refraction phenomenon can be exploited to form a novel kind of cavity, by blocking the outcoupling channel. This phenomenon has been predicted for 2D valleys~\cite{li2020}, but has not previously been verified experimentally. Using our 3D acoustic crystal, we realize such a cavity by placing a hard wall at the output facet (Fig.~\ref{fig4}G). The interface between the sample and the hard wall does not support propagating modes inside the bandgap (see the blue bands in Fig.~\ref{fig3}C). When the kink state arrives at the wall, it can only be reflected into other valleys; however, due to the suppression of intervalley scattering, the valley flipping process takes a long time, resulting in strong localization at the wall (Fig.~\ref{fig4}H). In Fig.~\ref{fig4}I, we plot the sound pressure amplitude along the propagation path for both open and closed boundaries at the output facet. For the open boundary case, the field decays as it propagates due to the intrinsic loss of the system. By contrast, under a closed boundary, the field is enhanced at the hard wall due to the cavity effect.

Finally, we demonstrate a functional device with 3D valley-dependent beam splitting by utilizing all six valleys to construct the kink states (Fig.~\ref{fig5}A). To this end, consider the (111) interface on which all valleys are separately projected, as illustrated in Fig.~\ref{fig5}B. Similar to the (001) interface, directional kink states with valley-locked group velocity arise at the valleys, which are found both numerically using the acoustic crystal (Fig.~\ref{fig5}C) and analytically using the effective Hamiltonians (See Supplementary text S4). Using these kink states at the six valleys, we conceive a structure as depicted in Fig.~\ref{fig5}A that consists of four interfaces, including one (111) interface and three other interfaces each normal to one of the Cartesian axes. The kink states on the (111) interface can couple to the ones on the other interfaces, splitting the six valley beams propagating on a 2D plane into separate beams propagating in 3D space, with their directions controlled by the valley indices. To experimentally probe this effect, we scan the acoustic fields at the outcoupling planes of each beam. As shown in Fig.~\ref{fig5}, D to F, each scanning plane has a single directional outcoupling beam as predicted.

\noindent\textbf{Discussion}

To conclude, we have demonstrated a 3D valley acoustic crystal with vectorial valley-contrasting physics, generalizing the valley concept from 2D to 3D.  The crystal hosts six valleys, grouped pairwise, each governed by an effective $2\times 2$ Dirac Hamiltonian involving two out of three momentum axes, with valley-contrasting Berry curvature and orbital magnetic moment. These 3D valleys give rise to valley vortex states in the bulk and valley kink states along 2D boundaries, allowing for novel forms of acoustic wave manipulation. The key properties of this acoustic crystal are captured by a two-band tight-binding model (Supplementary text S1), and should also be realizable in other systems such as photonic crystals~\cite{joannopoulos2008} and electric circuits~\cite{lee2018}. The orbital magnetic moments carried by the valley states may be useful for controlling the orbital angular momentum of light or sound waves, with potential applications for optical/acoustic tweezers~\cite{padgett2011, baresch2016} or communications channels~\cite{willner2015, shi2017}. The 3D valley kink states may be used in topological waveguides, cavities and antennas. Furthermore, with engineered non-Hermiticity and nonlinearity, the possibility of constructing 3D valley lasers and solitons can be explored~\cite{zeng2020, zhang2020}.  Finally, it would be highly interesting to find a real material realization, which could exhibit a 3D valley Hall effect (Supplementary text S5) and 3D valleytronics.

\noindent{\textbf{Materials and methods}}

\noindent\textbf{Numiercal simulation}

All numerical simulations are performed using the commercial software Comsol Multiphysics, pressure acoustics module. The bandstructures and eigenmodes (i.e., the results presented in Fig.~1, D to J, and Fig.~3, C and D, in the main text) are obtained from an eigen solver with periodic boundary conditions imposed on appropriate boundaries. The acoustic field distributions (i.e., the results presented in Fig.~4F in the main text) are obtained using a frequency domain solver with point sources operating at 8000 Hz. In all simulations, the density of air and sound speed are set as $\rho=1.3$ kg/$\text{m}^3$ and $c=343$ m/s, respectively. The largest allowed mesh size is smaller than one-tenth of the sound's wavelength in the air at 8000 Hz to ensure the accuracy of the finite-element calculations.

\noindent\textbf{Sample design and fabrication}

The detailed sample parameters are as follows: the side length of the cubic cell $a=31.5$ mm, the radii of the spherical cavities $r_1=6.8$ mm and $r_2=5.8$ mm (Hence $\Delta=1$ mm), and the diameter of the cylindrical tubes $d=8.1$ mm. The dimensions of the four samples used in the experiments are as follows: $8a\times 8a \times 8a$ (the sample used in Fig.~2), $14.75a\times 14.5a \times 9.5a$ (the sample used in Fig.~3 and Fig.~4, D to G), $9.75a\times 14.5a \times 9.5a$ (the sample used in Fig.~4, A to C) and $8a\times 8a \times 8a$ (Fig.~5, D to F). All samples are fabricated using the stereolithography technique, with a fabrication resolution of around 0.1 mm.

\noindent\textbf{Experimental measurement}

The sound signal is generated by a balanced armature speaker (E-audio type-31785). The excited acoustic field is scanned by a microphone (Br\"{u}el\&Kjær type-4961) mounted on a robotic arm. The measured signal is then processed by an analyzer (Br\"{u}el\&Kjær 3160-A-022 module) to obtain the amplitude and phase of sound for the frequencies of our interest (the frequency resolution is 2 Hz). To excite the bulk states (Fig.~2), the signal is guided to the center of the sample through a narrow tube. To excite the kink states at $\text{W}_2^+$ valley (Fig.~3 and Fig.~4), the speaker is directly placed on the wall of the sample at half height (i.e., at the location of the interface). To excite the kink states at all six valleys (Fig.~5), the signal is guided to the center of the (111) interface through a narrow tube.

\nocite{chang1996}
\nocite{cai2013}
\nocite{yang2021}
\nocite{fukui2005}
\nocite{xiao2010}
\nocite{wimmer2017}

\noindent \textbf{ACKNOWLEDGMENTS}

\noindent \textbf{Funding:} This work was supported by: Singapore
National Research Foundation Competitive Research Program (NRF-CRP23-2019-0005, NRF-CRP23-2019-0007 and NRF-CRP29-2022-0003) (to Y.C. and B.Z.), Singapore
National Research Foundation Investigatorship (NRF-NRFI08-2022-0001)(to Y.C.), Singapore Ministry of Education Academic Research Tier 2 grant (MOE-T2EP50123-0007)(to B.Z.), National Natural Science Foundation of China (12274183 and 12174159) (to H-X.S), National Key Research and Development Program of China (2020YFC1512403)(to H-X.S), and the Start-up Fund and the Direct Grant (4053675) of The Chinese University of Hong Kong (to H.X.). \textbf{Author contributions:} H.X. and Y.G. contributed equally to this work. H.X. conceived the idea. H.X., Z.C. and J.Z. performed the theoretical analysis and numerical calculation. Y.G., Y-J.G., H-Y. Z., S-Q.Y and H-X.S. conducted the experiment.  H.X., S.Y., Y.C. and B.Z. wrote the manuscript with input from all authors. H-X.S., Y.C. and B.Z. supervised the project.\textbf{Competing interests:} The authors declare no competing interests. \textbf{Data and materials availability:} All data needed to evaluate the conclusions in this paper are present in the paper and/or the Supplementary Materials.

\end{document}